\documentclass[11pt,a4paper]{article}%
\usepackage{amsfonts}
\usepackage{amsmath}
\usepackage{amssymb}
\usepackage{graphicx}
\usepackage{geometry}%
\providecommand{\U}[1]{\protect\rule{.1in}{.1in}}
\newtheorem{theorem}{Theorem}

\newtheorem{definition}{Definition}

\newtheorem{lemma}{Lemma}

\newtheorem{proposition}{Proposition}
\newtheorem{remark}{Remark}

\makeatletter
\def \@removefromreset#1#2{\let \@tempb \@elt
\def \@tempa#1{@&#1}\expandafter \let \csname @*#1*\endcsname \@tempa
\def \@elt##1{\expandafter \ifx \csname @*##1*\endcsname \@tempa \else
\noexpand \@elt{##1}\fi}     \expandafter \edef \csname cl@#2\endcsname{\csname cl@#2\endcsname}     \let \@elt \@tempb
\expandafter \let \csname @*#1*\endcsname \@undefined}

\@removefromreset{equation}{section}

\@removefromreset{theorem}{section}
\makeatother
\begin{document}

\title{The generalized Gell-Mann representation and violation of the CHSH inequality
by a general two-qudit state }
\author{Elena R. Loubenets\\National Research University Higher School of Economics, \\Moscow 101000, Russia}
\maketitle

\begin{abstract}
We formulate and prove the main properties of the generalized Gell-Mann
representation for traceless qudit observables with eigenvalues in $[-1,1]$
and analyze via this representation violation of the CHSH inequality by a
general two-qudit state. For the maximal value of the CHSH expectation in a
two-qudit state with an arbitrary qudit dimension $d\geq2$, this allows us to
find \emph{two new bounds, lower and upper}, expressed via the spectral
properties of the correlation matrix for a two-qudit state. We have not yet
been able to specify if the new upper bound improves the Tsirelson upper bound
for each two-qudit state. However, this is the case for all two-qubit states,
where the new lower bound and the new upper bound coincide and reduce to the
precise two-qubit CHSH\ result of Horodeckis, and also, for the
Greenberger--Horne--Zeilinger (GHZ) state with an odd $d\geq2,$ where the new
upper bound is less than the upper bound of Tsirelson. Moreover, we explicitly
find the correlation matrix for the two-qudit GHZ state and prove that, for
this state, the new upper bound is attained for each dimension $d\geq2$ and
this specifies the following \emph{new result: }for the two-qudit GHZ state,
the maximum of the CHSH\ expectation over traceless qudit observables with
eigenvalues in $[-1,1]$ is equal to $2\sqrt{2}$ if $d\geq2$ is even and to
$\frac{2(d-1)}{d}\sqrt{2}$ if $d>2$ is odd.

\end{abstract}

\section{Introduction}

Different aspects of quantum violation of the Clauser--Horne--Shimony--Holt
(CHSH) inequality \cite{1} were studied since 1969 in a huge number of papers
(see \cite{2,7,8,3,9,5,6} and references therein).

It is well known due to Tsirelson \cite{7,8} that, for all bipartite quantum
states, the maximal value of the CHSH expectation cannot exceed the
(Tsirelson) upper bound $2\sqrt{2}$ and that this upper bound is attained on
the two-qubit Bell states.

It is also well known that, for the maximum of the CHSH expectation in an
arbitrary two-qubit state over traceless qubit observables with eigenvalues
$\pm1$, Horodeckis \cite{9} found the precise value specified via the
correlation properties of this two-qubit state.

However, for a general two-qudit state with an arbitrary qudit dimension
$d>2,$ bounds on the maximal value of the CHSH expectation in terms of the
correlation properties of this state have not been analyzed.

In the present paper, we formulate and prove the main properties of the
generalized Gell-Mann representation for traceless qudit observables with
eigenvalues in $[-1,1]$ and study via this representation violation of the
CHSH inequality by a general two-qudit state.

For the maximal value of the CHSH expectation in a two-qudit state with an
arbitrary qudit dimension $d\geq2$, we find \emph{two} \emph{new bounds, lower
and upper, }expressed via the spectral properties of the correlation matrix of
this two-qudit state.

We have not yet been able to specify if the new upper bound improves the
Tsirelson upper bound for each two-qudit state. However, this is the case for
every two-qubit state, where the new lower bound and the new upper bound
coincide and reduce to the precise two-qubit result of Horodeckis in \cite{9},
and also, for the two-qudit Greenberger--Horne--Zeilinger (GHZ) state with an
arbitrary odd $d\geq2$, where the new upper bound is less\emph{\ }than the
upper bound of Tsirelson \cite{7,8}.

Applying our new general results for the two-qudit
Greenberger--Horne--Zeilinger (GHZ) state with an arbitrary $d\geq2$, we
explicitly find the correlation matrix for this state and prove that, for the
two-qudit GHZ state, the new upper bound is attained for each $d\geq2$ and
specifies the precise value of the CHSH\ expectation maximum in this state:
$2\sqrt{2}$ if $d\geq2$ is even and $\frac{2(d-1)}{d}\sqrt{2}$ if $d>2$ is odd.

The paper is organized as follows.

In Section 2, we formulate and prove (Theorem 1) the main properties of the
generalized Gell-Mann representation for traceless qudit observables with
eigenvalues in $[-1,1].$

In Section 3, via the properties of the generalized Gell-Mann representation
proved in Section 2, we study the maximal value of the CHSH expectation for a
general two-qudit state and derive new bounds (Theorem 2) on the maximal value
of the CHSH expectation for a two-qudit state with an arbitrary $d\geq2.$

In Section 4, we show that, for the GHZ state with an odd $d\geq2$, the new
upper bound is less (Proposition 2) than the upper bound of Tsirelson
\cite{7,8} and prove (Theorem 3) that, for the two-qudit GHZ state with an
arbitrary $d\geq2$, the new upper bound is attained.

In Section 5, we summarize the main new results of the present paper.

\section{Representation of traceless qudit observables}

For an arbitrary observable $Y$ on $\mathbb{C}^{d}$, $d\geq2,$ the generalized
Gell-Mann representation (see e. g. in \cite{10,11,12}) reads:%
\begin{align}
Y  &  =\alpha\mathbb{I}+n\cdot\Lambda,\text{ \ \ \ }n\cdot\Lambda
:=\sum\limits_{j=1,...,d^{2}-1}n_{j}\Lambda_{j},\label{1}\\
\alpha &  =\frac{1}{d}\mathrm{tr}[Y],\text{ \ \ \ }n_{j}=\frac{1}%
{2}\mathrm{tr}[Y\Lambda_{j}],\nonumber
\end{align}
where $n$ is a vector in $\mathbb{R}^{d^{2}-1}$ and $\Lambda_{j},$
$j=1,....,d^{2}-1,$ are traceless hermitian operators on $\mathbb{C}^{d}$
(generators of $SU(d)$ group), satisfying the relation%
\begin{equation}
\mathrm{tr}[\Lambda_{j}\Lambda_{j_{1}}]=2\delta_{jj_{1}}. \label{2}%
\end{equation}
The tuple $\Lambda:=(\Lambda_{1},...,\Lambda_{d^{2}-1})$ of these traceless
hermitian operators has the following form%
\begin{equation}
(\Lambda_{12}^{(s)},...,\Lambda_{1d}^{(s)},...,\Lambda_{d-1,d}^{(s)}%
,\Lambda_{12}^{(as)},...,\Lambda_{1d}^{(as)},...,\Lambda_{d-1,d}%
^{(as)},\Lambda_{1}^{(d)},...,\Lambda_{d-1}^{(d)}), \label{3}%
\end{equation}
where%
\begin{align}
\Lambda_{mk}^{(s)}  &  =\left\vert m\right\rangle \left\langle k\right\vert
+\left\vert k\right\rangle \left\langle m\right\vert ,\text{ \ \ \ \ }%
\Lambda_{mk}^{s}=\Lambda_{km}^{s},\text{ \ \ \ }1\leq m<k\leq d,\label{4.1}\\
& \nonumber\\
\Lambda_{mk}^{(as)}  &  =-i\text{ }\left\vert m\right\rangle \left\langle
k\right\vert +i\text{ }\left\vert k\right\rangle \left\langle m\right\vert
,\text{ \ \ }\Lambda_{mk}^{as}=-\Lambda_{km}^{as},\text{ \ \ \ }1\leq m<k\leq
d\label{4.2}\\
& \nonumber\\
\Lambda_{l}^{(d)}  &  =\sqrt{\frac{2}{l(l+1)}}\left(  \sum\limits_{j=1,...,l}%
\left\vert j\right\rangle \left\langle j\right\vert -l\text{ }\left\vert
l+1\right\rangle \left\langle l+1\right\vert \right)  ,\text{ \ \ }1\leq l\leq
d-1, \label{4.3}%
\end{align}
and $\{\left\vert j\right\rangle \in\mathbb{C}^{d},j=1,...,d\}$ is the
computational basis of $\mathbb{C}^{d}.$ The matrix representations of
operators $\Lambda_{i},$ $i=1,....,d^{2}-1$ in the computational basis of
$\mathbb{C}^{d}$ constitute the higher-dimensional extensions of the Pauli
matrices for qubits ($d=2$) and the Gell-Mann matrices for qutrits ($d=3$).

Representation (\ref{1}) constitutes decomposition of a qudit observable $Y$
in the orthonormal basis $\{\mathbb{I},\Lambda_{j},j=1,....,d^{2}-1\}$ of the
Hilbert-Schmidt space $\mathcal{H}_{sch}$ where qudit observables are vectors
and the scalar product is defined via $\langle Y,Y^{\prime}\rangle
_{\mathcal{H}_{sch}}:=\mathrm{tr}[YY^{\prime}].$

Note that, for derivation of our new results in Sections 2, 3, a choice in
(\ref{1}) of a basis of $\mathcal{H}_{sch}$ is not essential -- we could take
any basis of $\mathcal{H}_{sch}.$ Though the new result presented in Section 4
also does not itself depend on a basis choice in representation (\ref{1}), our
proof of this result is based specifically on basis (\ref{3}).

For a traceless qudit observable $X$, representation (\ref{1}) takes the form
\begin{equation}
X=n\cdot\Lambda,\text{ \ \ \ }n_{j}=\frac{1}{2}\mathrm{tr}[X\Lambda_{j}],
\label{5}%
\end{equation}
and implies
\begin{equation}
\mathrm{tr}[X^{2}]=2\left\Vert n\right\Vert ^{2}. \label{6}%
\end{equation}
The following statement is proved in Appendix A.

\begin{lemma}
For each $n\in\mathbb{R}^{d^{2}-1}$ and each $d\geq2,$
\begin{equation}
\sqrt{\frac{2}{d}}\text{ }\leq\frac{\left\Vert n\cdot\Lambda\right\Vert _{0}%
}{\left\Vert n\right\Vert }\text{ }\leq\text{ }\sqrt{\frac{2(d-1)}{d}}\text{
}\leq\sqrt{\frac{2}{1+\delta_{d2}}}, \label{7}%
\end{equation}
where (i) notation $\left\Vert \cdot\right\Vert _{0}$ means the operator norm
of an observable on $\mathbb{C}^{d}$\ and $\left\Vert \cdot\right\Vert $ --
the Euclidian norm of a vector $n$ in $\mathbb{R}^{d^{2}-1};$ (ii)
$\delta_{d2}=1$ for $d=2$ and $\delta_{d2}=0,$ otherwise.
\end{lemma}

From the lower bound in (\ref{7}) it follows that, for each $n\in
\mathbb{R}^{d^{2}-1},$ relation $\left\Vert n\cdot\Lambda\right\Vert _{0}%
\leq\sqrt{\frac{2}{d}}$ implies $\left\Vert n\right\Vert \leq1,$ and if
$\left\Vert n\cdot\Lambda\right\Vert _{0}\leq\sqrt{\frac{2}{d}}\left\Vert
n\right\Vert ,$ then \ $\left\Vert n\cdot\Lambda\right\Vert _{0}=\sqrt
{\frac{2}{d}}\left\Vert n\right\Vert $.

\begin{remark}
In the qubit case ($d=2)$, operators $\Lambda_{j},$ $j=1,2,3,$ constitute the
Pauli operators $\sigma_{j},j=1,2,3$, and bounds (\ref{7}) reduce to the
well-known relation
\begin{equation}
\frac{\left\Vert n\cdot\sigma\right\Vert _{0}}{\left\Vert n\right\Vert
}{\Large \mid}_{d=2}=1,\text{ \ \ \ }n\in\mathbb{R}^{3}, \label{8}%
\end{equation}
valid for any traceless qubit observable $n\cdot\sigma$ -- projection of the
qubit spin on a direction $n$ in $\mathbb{R}^{3}.$
\end{remark}

In what follows, we use the following notations, for short.

\begin{definition}
Denote by $\mathcal{L}_{d}$ the set of all traceless qudit observables on
$\mathbb{C}^{d}$ with eigenvalues in $[-1,1]$ and by $\mathcal{L}_{d}%
^{(s)}\subset\mathcal{L}_{d}$ the subset of all traceless qudit observables
$X_{s}$ with a dimension $s\geq0$ of their kernels and all their eigenvalues
in set $\{-1,0,1\}$. Subset $\mathcal{L}_{d}^{(0)}\neq\varnothing$ iff a
dimension $d\geq2$ is even.
\end{definition}

If $s>0$, then it constitutes the multiplicity of the zero eigenvalue of an
observable in $\mathcal{L}_{d}^{(s)}$. If $s=0,$ then an observable $X_{0}%
\in\mathcal{L}_{d}^{(0)}$ does not have the zero eigenvalue.

Normalizing a vector $n$ in decomposition (\ref{5}) in view of the lower bound
in (\ref{7}), we represent each traceless qudit observable $X\in
\mathcal{L}_{d}$ in the form%
\begin{align}
X  &  =\sqrt{\frac{d}{2}}\text{ }\left(  n_{X}\cdot\Lambda\right)  ,\text{
\ \ }n_{X}^{(j)}=\frac{1}{\sqrt{2d}}\text{ }\mathrm{tr}[X\Lambda_{j}],\text{
\ \ }n_{X}\in\mathbb{R}^{d^{2}-1},\label{9}\\
\mathrm{tr}[X^{2}]  &  =d\left\Vert n_{X}\right\Vert ^{2}, \label{10}%
\end{align}
so that, under representation (\ref{9}),
\begin{equation}
\left\Vert X\right\Vert _{0}\leq1\text{ \ }\Leftrightarrow\text{ }\left\Vert
n_{X}\cdot\Lambda\right\Vert _{0}\leq\sqrt{\frac{2}{d}}. \label{11}%
\end{equation}
Therefore, representation (\ref{9}) establishes the mapping
\begin{equation}
\Phi:\mathcal{L}_{d}\rightarrow\mathfrak{R}_{d} \label{12}%
\end{equation}
of all traceless observables in $\mathcal{L}_{d}$ to vectors $n\in
\mathbb{R}^{d^{2}-1}$ in the set
\begin{equation}
\mathfrak{R}_{d}=\left\{  n\in\mathbb{R}^{d^{2}-1}\mid\left\Vert n\cdot
\Lambda\right\Vert _{0}\leq\sqrt{\frac{2}{d}}\right\}  \label{13}%
\end{equation}
and since (\ref{9}) constitutes the decomposition via the basis $\{\mathbb{I}%
,$ $\Lambda_{j},$ $j=1,....,d^{2}-1\}$ of the Hilbert-Schmidt space
$\mathcal{H}_{sch}$, the mapping $\Phi$ is \emph{injective}.

Under representation (\ref{9}), for all observables $X\in\mathcal{L}_{d}%
^{(s)}$,
\begin{equation}
\mathrm{tr}[X^{2}]=(d-s)=d\left\Vert n_{X}\right\Vert ^{2}. \label{14}%
\end{equation}
Therefore, all traceless observables in $\mathcal{L}_{d}^{(s)}\subset
\mathcal{L}_{d},$ $s\geq0,$ are mapped to vectors $n$ in
\begin{equation}
\mathfrak{R}_{d}^{(s)}=\left\{  n\in\mathbb{R}^{d^{2}-1}\mid\left\Vert
n\cdot\Lambda\right\Vert _{0}=\sqrt{\frac{2}{d}},\text{ \ }\left\Vert
n\right\Vert =\sqrt{\frac{d-s}{d}}\right\}  \subset\mathfrak{R}_{d}%
,\text{\ \ }s\geq0. \label{15}%
\end{equation}
In particular, for an even $d\geq2,$ all traceless qudit observables with
eigenvalues $\pm1$ (i. e. in subset $\mathcal{L}_{d}^{(0)}\subset
\mathcal{L}_{d}),$ are mapped to vectors in
\begin{equation}
\mathfrak{R}_{d}^{(0)}=\left\{  n\in\mathbb{R}^{d^{2}-1}\mid\left\Vert
n\cdot\Lambda\right\Vert _{0}=\sqrt{\frac{2}{d}},\text{ \ }\left\Vert
n\right\Vert =1\right\}  \subset\mathfrak{R}_{d}. \label{16}%
\end{equation}

Conversely, let $n$ be an arbitrary vector in set $\mathfrak{R}_{d},$ defined
by (\ref{13}). For the traceless qudit observable, corresponding to each
$n\in\mathfrak{R}_{d}$ via the representation
\begin{equation}
X_{n}=\sqrt{\frac{d}{2}}\text{ }\left(  n\cdot\Lambda\right)  ,\text{
\ \ \ \ }\mathrm{tr}[X_{n}^{2}]=d\left\Vert n\right\Vert ^{2}, \label{17}%
\end{equation}
the operator norm $\left\Vert X_{n}\right\Vert _{0}\leq1$ and, therefore,
$X_{n}\in\mathcal{L}_{d}$. \ Hence, \emph{the injective mapping (\ref{12}) is
surjective, therefore,} \emph{bijective}.

Furthermore, let subset $\mathfrak{R}_{d}^{(0)}\subset\mathfrak{R}_{d}$, given
by (\ref{16}), be not empty and $r\in\mathfrak{R}_{d}^{(0)}$ be a unit vector
in this subset. For the observable $X_{r}=\sqrt{\frac{d}{2}}\left(
r\cdot\Lambda\right)  $ in $\mathcal{L}_{d}$ corresponding to a unit vector
$r\in\mathfrak{R}_{d}^{(0)}$ via (\ref{17}), we have
\begin{equation}
\left\Vert X_{r}\right\Vert _{0}=1\text{, \ \ \ \ \ }\mathrm{tr}[X_{r}^{2}]=d.
\label{18}%
\end{equation}
But this is possible iff $X_{r}\in\mathcal{L}_{d}^{(0)}$. Therefore,
representation (\ref{17}) establishes the one-to-one correspondence between
observables in the subset $\mathcal{L}_{d}^{(0)}\subset\mathcal{L}_{d}$ and
vectors in the intersection $\mathfrak{R}_{d}^{(0)}$ of $\mathfrak{R}_{d}$
with the unit sphere. Since $\mathcal{L}_{d}^{(0)}\neq\varnothing$ iff a
dimension $d\geq2$ is even, also, subset $\mathfrak{R}_{d}^{(0)}%
\neq\varnothing$ iff a dimension $d\geq2$ is even.

Let us now analyze \emph{geometry }of the set $\mathfrak{R}_{d}$, $d\geq2,$
defined by (\ref{13}).

As specified in Remark 1, in the qubit ($d=2$) case, for each $n\in
\mathbb{R}^{3},$ the norm $\left\Vert n\cdot\sigma\right\Vert _{0}=\left\Vert
n\right\Vert $. Therefore, in (\ref{13}), relation $\left\Vert n\cdot
\Lambda\right\Vert _{0}\leq1$ is equivalent to $\left\Vert n\right\Vert
\leq1,$ the set
\begin{equation}
\mathfrak{R}_{2}=\{n\in\mathbb{R}^{3}\mid\left\Vert n\right\Vert \leq1\}
\label{18.2}%
\end{equation}
constitutes the unit ball in $\mathbb{R}^{3}$.

Due to (\ref{13}), for each vector $n\in\mathfrak{R}_{d}$ along $j$-$th$
coordinate axis in $\mathbb{R}^{d^{2}-1}$, the norm cannot exceed
\begin{equation}
\sqrt{\frac{2}{d}}\text{ }\frac{1}{\left\Vert \Lambda_{j}\right\Vert _{op}%
},\text{ \ \ \ \ }j=1,...,d^{2}-1, \label{19}%
\end{equation}
where the operator norms of all observables (\ref{4.1}) and (\ref{4.2}) are
equal to $1$, whereas the operator norms of observables (\ref{4.3}) vary from
$1$ to $\sqrt{\frac{2(d-1)}{d}}.$ Therefore, the lengths (\ref{19}) of vectors
along different coordinate axes in $\mathbb{R}^{d^{2}-1}$ vary from
$\sqrt{\frac{2}{d}}$ to $\sqrt{\frac{1}{d-1}}$ and are equal to each other if
only $d=2.$ Thus, for $d>2,$ the form of the bounded set $\mathfrak{R}_{d}$
with respect to different axes is asymmetric.

The maximal norm $l_{d}:=\max_{n\in\mathfrak{R}_{d}}\left\Vert n\right\Vert $
of a vector $n\in\mathfrak{R}_{d}$ is calculated via (\ref{10}):
\begin{equation}
l_{d}:=\max_{n\in\mathfrak{R}_{d}}\left\Vert n\right\Vert =\sqrt{\frac{1}%
{d}\max_{X\in\mathcal{L}_{d}}\mathrm{tr}[X^{2}]}=\sqrt{\frac{1}{d}%
\max_{\lambda_{m}\in D}%
{\textstyle\sum\limits_{m}}
\lambda_{m}^{2},} \label{20}%
\end{equation}
and is equal to
\begin{align}
l_{d}  &  =1,\ \ \ \ \ \ \ \ \ \ \ \ \text{if }d\geq2\text{ \ is
\ even,}\label{21}\\
l_{d}  &  =\sqrt{\frac{d-1}{d}},\text{ \ \ if }d\geq2\text{ \ is
\ odd.}\nonumber
\end{align}
In (\ref{20}), $\{\lambda_{m}\in\lbrack-1,1],$ $m=1,...,d\}$ are eigenvalues
of a traceless observable $X=\sqrt{\frac{d}{2}}\left(  n\cdot\Lambda\right)
\in\mathcal{L}_{d}$ and $D=\left\{  \left\vert \lambda_{m}\right\vert
\leq1,\text{ }m=1,...,d,\text{ }\sum_{m}\lambda_{m}=0\right\}  .$

For an even $d\geq2,\ $the maximal norm $l_{d}$ is attained on vectors
$n\in\mathfrak{R}_{d}^{(0)},$ corresponding to observables in $\mathcal{L}%
_{d}^{(0)},$ for an odd $d\geq2\ -$ on vectors $n\in\mathfrak{R}_{d}^{(1)}$,
corresponding to observables in $\mathcal{L}_{d}^{(1)}$ (see Definition 1).

From (\ref{20}) it follows, that, for all $n\in\mathfrak{R}_{d},$ the norms
\begin{equation}
\left\Vert n\right\Vert \leq l_{d}\leq1, \label{22}%
\end{equation}
so that $\mathfrak{R}_{d}$ is a subset of the ball in $\mathbb{R}^{d^{2}-1}$of
radius $l_{d}\leq1.$ Relations (\ref{7}), (\ref{13}) imply that%
\begin{equation}
\mathfrak{R}_{d}\supseteq\left\{  n\in\mathbb{R}^{d^{2}-1}\mid\left\Vert
n\right\Vert =\sqrt{\frac{1}{d-1}}\right\}  , \label{23}%
\end{equation}
therefore, $\mathfrak{R}_{d}$ contains the ball of radius $\sqrt{\frac{1}%
{d-1}}.$ Note also that, for each vector $n\in\mathbb{R}^{d^{2}-1},$ the
vector
\begin{equation}
\widetilde{n}=\sqrt{\frac{2}{d}}\text{ }\frac{n}{\left\Vert n\cdot
\Lambda\right\Vert _{0}}\in\mathfrak{R}_{d}. \label{23.1}%
\end{equation}

Summing up, we have proved the following statement.

\begin{theorem}
Representation
\begin{equation}
X=\sqrt{\frac{d}{2}}\text{ }\left(  n\cdot\Lambda\right)  \label{24}%
\end{equation}
establishes the one-to-one correspondence $\mathcal{L}_{d}\leftrightarrow
\mathfrak{R}_{d}$ between traceless qudit observables $X$ with eigenvalues in
$[-1,1]$ and vectors $n$ in the set $\mathfrak{R}_{d},$ defined by (\ref{13})
and having, in general, the complicated form specified above by (\ref{18.2}%
)--( \ref{23.1}). The maximal norm $l_{d}$ of a vector in $\mathfrak{R}_{d}$
is equal to $l_{d}$ $=1$ if a qudit dimension $d\geq2$ is even and to $l_{d}$
$=\sqrt{\frac{d-1}{d}}$ if a qudit dimension $d\geq2$ is odd. Under the
one-to-one correspondence $\mathcal{L}_{d}\leftrightarrow\mathfrak{R}_{d}$
established by (\ref{24}), sets%
\begin{equation}
\mathcal{L}_{d}^{(0)}\leftrightarrow\mathfrak{R}_{d}^{(0)}=\left\{
\ n\in\mathfrak{R}_{d}\mid\left\Vert n\right\Vert =1\right\}  \label{25}%
\end{equation}
and are not empty iff a dimension $d\geq2$ is even.
\end{theorem}

In the next section, we use Theorem 1 for analyzing violation of the CHSH
inequality by an arbitrary two-qudit state.

\section{The maximal value of the CHSH expectation for a two-qudit state}

Let $\rho_{d\times d}$ be a state on $\mathbb{C}^{d}\otimes\mathbb{C}^{d}$ and
$A_{i},B_{k},$ $i,k=1,2$, be traceless observables on $\mathbb{C}^{d}$ with
eigenvalues in $[-1,1],$ that is, $A_{i},B_{k}\in\mathcal{L}_{d}$ (see
Definition 1).

Consider a bipartite correlation scenario\footnote{On the general framework
for the probabilistic description of a multipartite correlation scenario with
arbitrary numbers of settings and outcomes per site, see \cite{13.1}.} where
each of two parties, say, Alice and Bob, performs measurements of two qudit
observables in a state $\rho_{d\times d}$. Let Alice measure observables
$A_{1},A_{2}\in\mathcal{L}_{d}$ and Bob -- observables $B_{1},B_{2}%
\in\mathcal{L}_{d}$. For this quantum bipartite scenario, the Bell operator
associated with the CHSH inequality has the form
\begin{equation}
\mathcal{B}_{chsh}(A_{1},A_{2};B_{1},B_{2})=A_{1}\otimes(B_{1}+B_{2}%
)+A_{2}\otimes(B_{1}-B_{2}) \label{34}%
\end{equation}
and, for a state $\rho_{d\times d},$ the left-hand side of the CHSH
inequality\footnote{In the local hidden variable (LHV) model, the expectation
$\left\vert \left\langle \mathcal{B}_{chsh}\right\rangle _{lhv}\right\vert
\leq2$ -- the CHSH inequality \cite{1}.} reads:%
\begin{align}
&  \left\vert \mathrm{tr}[\rho_{d\times d}\mathcal{B}_{chsh}(A_{1},A_{2}%
;B_{1},B_{2})]\right\vert \label{35}\\
&  =\left\vert \text{ }\mathrm{tr}[\rho_{d\times d}\{A_{1}\otimes(B_{1}%
+B_{2})\}]+\mathrm{tr}[\rho_{d\times d}\{A_{2}\otimes(B_{1}-B_{2})\}]\text{
}\right\vert .\nonumber
\end{align}

\begin{remark}
For a bipartite correlation scenario with two observables per site, the CHSH
inequality was originally \cite{1} introduced and proved in case of dichotomic
observables with outcomes $\pm1$ at each site. It was, however, further proved
that, in the LHV frame, the CHSH inequality holds for a bipartite correlation
scenario with outcomes in $[-1,1]$ of any spectral type. For the proof, see,
for example, Proposition 1.4.3 in \cite{13.2} and also, Section 3.2 in
\cite{13.3}. Note also that by Corollaries 1, 2 in \cite{13.3} each of known
correlation Bell inequalities, introduced originally for scenarios with
outcomes $\pm1,$ holds also for scenarios with outcomes in $[-1,1]$ of any
spectral type.
\end{remark}

For the expectation (\ref{35}) of the Bell operator (\ref{34}) in a state
$\rho_{d\times d},$ let us analyze the least upper bound
\begin{equation}
\sup_{A_{i},B_{k}\in\mathcal{L}_{d}}\left\vert \text{ }\mathrm{tr}%
[\rho_{d\times d}\mathcal{B}_{chsh}(A_{1},A_{2};B_{1},B_{2})]\right\vert
\label{35_}%
\end{equation}
over observables $A_{i},B_{k}\in\mathcal{L}_{d}.$

For these observables, representation (\ref{24}) reads
\begin{equation}
A_{i}=\sqrt{\frac{d}{2}}\text{ }\left(  a_{i}\cdot\Lambda\right)  ,\text{
\ \ }B_{k}=\sqrt{\frac{d}{2}}\text{ }\left(  b_{k}\cdot\Lambda\right)
,\text{\ \ \ }a_{i},b_{k}\in\mathfrak{R}_{d}\subset\mathbb{R}^{d^{2}-1},
\label{33}%
\end{equation}
where set $\mathfrak{R}_{d}$ is given by (\ref{13}) and specified in Theorem
1. By (\ref{33}) each expectation $\mathrm{tr}[\rho_{d\times d}\left\{
A_{i}\otimes B_{k}\right\}  ]$ in (\ref{35}) takes the form%
\begin{align}
\mathrm{tr}[\rho_{d\times d}\left\{  A_{m}\otimes B_{k}\right\}  ]  &
=\frac{d}{2}%
{\textstyle\sum\limits_{n,m}}
a_{m}^{(i)}T_{\rho_{d\times d}}^{(ij)}b_{k}^{(j)}\label{36}\\
&  =\frac{d}{2}\left\langle a_{m},T_{\rho_{d\times d}}b_{k}\right\rangle
,\nonumber
\end{align}
where $T_{\rho_{d\times d}}$ is the linear operator on $\mathbb{R}^{d^{2}-1}$,
defined in the standard basis of $\mathbb{R}^{d^{2}-1}$ via the two-qudit
correlation matrix with real elements:
\begin{equation}
T_{\rho_{d\times d}}^{(ij)}:=\mathrm{tr}[\rho_{d\times d}\{\Lambda_{i}%
\otimes\Lambda_{j}\}],\text{ \ }i,j=1,...,d^{2}-1. \label{37}%
\end{equation}
This $\left(  d^{2}-1\right)  \times\left(  d^{2}-1\right)  $ matrix
constitutes a generalization to higher dimensions of the two-qubit correlation
matrix, considered in \cite{9,12}. If a state $\rho_{d\times d}$ is invariant
under permutation of spaces $\mathbb{C}^{d}$ in the tensor product
$\mathbb{C}^{d}\otimes\mathbb{C}^{d}$, \emph{permutationally invariant}, for
short, then $T_{\rho_{d\times d}}$ is hermitian. Note that since $\left\vert
\text{ }\mathrm{tr}[\rho_{d\times d}\left\{  A\otimes B\right\}  ]\text{
}\right\vert \leq1$ for any state $\rho_{d\times d}$ and all observables
$A,B\in\mathcal{L}_{d},$ relation (\ref{36}) implies
\begin{equation}
\left\vert \left\langle a,T_{\rho_{d\times d}}b\right\rangle \right\vert
\leq\frac{2}{d},\text{ \ \ for all }a,b\in\mathfrak{R}_{d}. \label{37-1}%
\end{equation}

Substituting (\ref{36}) into (\ref{35}), we have%
\begin{equation}
\mathrm{tr}[\rho_{d\times d}\mathcal{B}_{chsh}(A_{1},A_{2};B_{1},B_{2}%
)]=\frac{d}{2}\left\{  \text{ }\left\langle a_{1},T_{\rho_{d\times d}}%
(b_{1}+b_{2})\right\rangle +\left\langle a_{2},T_{\rho_{d\times d}}%
(b_{1}-b_{2})\right\rangle \right\}  . \label{38}%
\end{equation}
Due to the one-to-one correspondence $\mathcal{L}_{d}\leftrightarrow
\mathfrak{R}_{d}$ (see Theorem 1) between traceless qudit observables with
eigenvalues in $[-1,1]$ and vectors in the subset $\mathfrak{R}_{d}$ of
$\mathbb{R}^{d^{2}-1}$ given by (\ref{13}), relation (\ref{38}) implies%
\begin{align}
&  \sup_{A_{i},B_{k}\in\mathcal{L}_{d}}\left\vert \mathrm{tr}[\rho_{d\times
d}\mathcal{B}_{chsh\ }(A_{1},A_{2};B_{1},B_{2})]\right\vert \label{39}\\
&  =\frac{d}{2}\sup_{a_{i},b_{k}\in\mathfrak{R}_{d}}\left\vert \text{
}\left\langle a_{1},T_{\rho_{d\times d}}(b_{1}+b_{2})\right\rangle
+\left\langle a_{2},T_{\rho_{d\times d}}(b_{1}-b_{2})\right\rangle \right\vert
.\nonumber
\end{align}
Taking into account continuity in $a_{i},b_{k}$ of scalar products on
$\mathbb{R}^{d^{2}-1},$ boundedness and closure of set $\mathfrak{R}_{d}$ and
the one-to-one correspondence $\mathcal{L}_{d}\leftrightarrow\mathfrak{R}%
_{d},$ we have
\begin{align}
&  \sup_{A_{i},B_{k}\in\mathcal{L}_{d}}\left\vert \mathrm{tr}[\rho_{d\times
d}\mathcal{B}_{chsh\ }(A_{1},A_{2};B_{1},B_{2})]\right\vert \label{40}\\
& \nonumber\\
&  =\max_{A_{i},B_{k}\in\mathcal{L}_{d}}\left\vert \mathrm{tr}[\rho_{d\times
d}\mathcal{B}_{chsh\ }(A_{1},A_{2};B_{1},B_{2})]\right\vert \nonumber\\
& \nonumber\\
&  =\frac{d}{2}\max_{a_{i},b_{k}\in\mathfrak{R}_{d}}\text{ }\left\vert \text{
}\left\langle a_{1},T_{\rho_{d\times d}}(b_{1}+b_{2})\right\rangle
+\left\langle a_{2},T_{\rho_{d\times d}}(b_{1}-b_{2})\right\rangle \right\vert
.\nonumber
\end{align}
In the last line of (\ref{40}), the maximum over $a_{i}\in\mathfrak{R}_{d},$
$i=1,2,$ is attained on vectors
\begin{equation}
\widetilde{a}_{1}=\sqrt{\frac{2}{d}}\text{ }\frac{T_{\rho_{d\times d}}%
(b_{1}+b_{2})}{\left\Vert T_{\rho_{d\times d}}(b_{1}+b_{2})\cdot
\Lambda\right\Vert _{0}}\in\mathfrak{R}_{d},\text{ \ \ \ \ }\widetilde{a}%
_{2}=\sqrt{\frac{2}{d}}\text{ }\frac{T_{\rho_{d\times d}}(b_{1}-b_{2}%
)}{\left\Vert T_{\rho_{d\times d}}(b_{1}-b_{2})\cdot\Lambda\right\Vert _{0}%
}\in\mathfrak{R}_{d}, \label{41}%
\end{equation}
which are not necessarily unit, and is equal to%
\begin{equation}
\sqrt{\frac{d}{2}}\left(  \frac{\left\Vert T_{\rho_{d\times d}}(b_{1}%
+b_{2})\right\Vert ^{2}}{\left\Vert T_{\rho_{d\times d}}(b_{1}+b_{2}%
)\cdot\Lambda\right\Vert _{0}}+\frac{\left\Vert T_{\rho_{d\times d}}%
(b_{1}-b_{2})\right\Vert ^{2}}{\left\Vert T_{\rho_{d\times d}}(b_{1}%
-b_{2})\cdot\Lambda\right\Vert _{0}}\right)  . \label{42}%
\end{equation}
Noting further that by (\ref{13}), for arbitrary $b_{1},b_{2}\in
\mathfrak{R}_{d},$ vectors\ $\frac{b_{1}\pm b_{2}}{2}$ are also in
$\mathfrak{R}_{d}$, we generalize the calculation method used by Horodeckis in
\cite{9} and introduce two vectors $r_{1},r_{2}\in\mathfrak{R}_{d}$, not
necessarily mutually orthogonal and satisfying the relations
\begin{align}
\frac{b_{1}+b_{2}}{2}  &  =r_{1}\cos\theta,\text{\ \ \ \ }\frac{b_{1}-b_{2}%
}{2}=r_{2}\sin\theta,\text{ \ }\theta\in\lbrack0,\pi/2],\label{43}\\
2\left\langle r_{1},r_{2}\right\rangle \sin2\theta &  =\left\Vert
b_{1}\right\Vert ^{2}-\left\Vert b_{2}\right\Vert ^{2}.\nonumber
\end{align}
Substituting (\ref{43}) into (\ref{42}) and (\ref{42}) into (\ref{40}), we
have
\begin{align}
&  \frac{d}{2}\max_{a_{i},b_{k}\in\mathfrak{R}_{d}}\left\vert \text{
}\left\langle a_{1},T_{\rho_{d\times d}}(b_{1}+b_{2})\right\rangle
+\left\langle a_{2},T_{\rho_{d\times d}}(b_{1}-b_{2})\right\rangle \text{
}\right\vert \label{44}\\
& \nonumber\\
&  =\sqrt{2d}\max_{\theta,\text{ }r_{1,}r_{2}\in\mathfrak{R}_{d}}\left(
\frac{\left\Vert T_{\rho_{d\times d}}r_{1}\right\Vert ^{2}}{\left\Vert
T_{\rho_{d\times d}}r_{1}\cdot\Lambda\right\Vert _{0}}\cos\theta
+\frac{\left\Vert T_{\rho_{d\times d}}r_{2}\right\Vert ^{2}}{\left\Vert
T_{\rho_{d\times d}}r_{2}\cdot\Lambda\right\Vert _{0}}\sin\theta\right)
,\nonumber
\end{align}
where the maximum over $\theta\in\lbrack0,\pi/2]$\ is attained at%
\begin{equation}
\mathrm{tg}\text{ }\theta_{0}=\frac{\left\Vert T_{\rho_{d\times d}}%
r_{2}\right\Vert ^{2}}{\left\Vert T_{\rho_{d\times d}}r_{1}\right\Vert ^{2}%
}\cdot\frac{\left\Vert T_{\rho_{d\times d}}r_{1}\cdot\Lambda\right\Vert _{0}%
}{\left\Vert T_{\rho_{d\times d}}r_{2}\cdot\Lambda\right\Vert _{0}} \label{45}%
\end{equation}
and is given by%
\begin{equation}
\sqrt{\frac{\left\Vert T_{\rho_{d\times d}}r_{1}\right\Vert ^{4}}{\left\Vert
T_{\rho_{d\times d}}r_{1}\cdot\Lambda\right\Vert _{0}^{2}}+\frac{\left\Vert
T_{\rho_{d\times d}}r_{2}\right\Vert ^{4}}{\left\Vert T_{\rho_{d\times d}%
}r_{2}\cdot\Lambda\right\Vert _{0}^{2}}}. \label{46}%
\end{equation}

From (\ref{40})--(\ref{46}) it follows
\begin{align}
&  \max_{A_{i},B_{k}\in\mathcal{L}_{d}}\left\vert \mathrm{tr}[\rho_{d\times
d}\mathcal{B}_{chsh\ }(A_{1},A_{2};B_{1},B_{2})]\right\vert \label{47}\\
& \nonumber\\
&  =\sqrt{2d}\max_{r_{1,}r_{2}\in\mathfrak{R}_{d}}\sqrt{\frac{\left\Vert
T_{\rho_{d\times d}}r_{1}\right\Vert ^{4}}{\left\Vert T_{\rho_{d\times d}%
}r_{1}\cdot\Lambda\right\Vert _{0}^{2}}+\frac{\left\Vert T_{\rho_{d\times d}%
}r_{2}\right\Vert ^{4}}{\left\Vert T_{\rho_{d\times d}}r_{2}\cdot
\Lambda\right\Vert _{0}^{2}}}.\nonumber
\end{align}

Let us first show that the derived new expression (\ref{47}) for maximum
(\ref{40}) leads at once to the Tsirelson upper bound. Namely, due to property
(\ref{23.1}) and relation (\ref{37-1}), specified with vectors%
\begin{align}
a  &  =\sqrt{\frac{2}{d}}\text{ }\frac{T_{\rho_{d\times d}}r}{\left\Vert
T_{\rho_{d\times d}}r\cdot\Lambda\right\Vert _{0}}\in\mathfrak{R}%
_{d},\label{48.1}\\
b  &  =r\in\mathfrak{R}_{d},\nonumber
\end{align}
we have the bound
\begin{equation}
\frac{\left\Vert T_{\rho_{d\times d}}r\right\Vert ^{2}}{\left\Vert
T_{\rho_{d\times d}}r\cdot\Lambda\right\Vert _{0}}\text{ }\leq\text{ }%
\sqrt{\frac{2}{d}},\text{ \ \ for all }r\in\mathfrak{R}_{d}. \label{48}%
\end{equation}
Substituting (\ref{48}) into (\ref{47}), we have%
\begin{equation}
\max_{A_{i},B_{k}\in\mathcal{L}_{d}}\left\vert \mathrm{tr}[\rho_{d\times
d}\mathcal{B}_{chsh\ }(A_{1},A_{2};B_{1},B_{2})]\right\vert \text{ }\leq\text{
}2\sqrt{2},\text{ \ \ \ \ \ \ \ }\forall\rho_{d\times d},\text{ }\forall
d\geq2, \label{49}%
\end{equation}
that is, the Tsirelson upper bound derived originally \cite{7} in the other way.

Furthermore, the derived expression (\ref{47}) for the maximal value of
(\ref{35}) allows us to introduce the following two \emph{new} \emph{general
bounds, lower and upper,} see Appendix B for the proof.\emph{\ }

\begin{theorem}
For an arbitrary two-qudit state $\rho_{d\times d},$ $d\geq2,$ with the
correlation matrix $T_{\rho_{d\times d}}$ defined by (\ref{37}), the maximal
value of the CHSH expectation (\ref{35}) over all traceless qudit observables
with eigenvalues in $[-1,1]$ admits the bounds%
\begin{align}
&  \frac{d}{d-1}\sqrt{\lambda_{\rho_{d\times d}}+\widetilde{\lambda}%
_{\rho_{d\times d}}}\label{51}\\
& \nonumber\\
&  \leq\max_{A_{i},B_{k}\in\mathcal{L}_{d}}\text{ }\left\vert \mathrm{tr}%
[\rho_{d\times d}\mathcal{B}_{chsh}(A_{1},A_{2};B_{1},B_{2})]\right\vert
\nonumber\\
& \nonumber\\
&  \leq l_{d}^{2}d\sqrt{\lambda_{\rho_{d\times d}}+\widetilde{\lambda}%
_{\rho_{d\times d}}},\nonumber
\end{align}
where $\lambda_{\rho_{d\times d}}\geq$ $\widetilde{\lambda}_{\rho_{d\times d}%
}\geq0$\textbf{\ }are two greater eigenvalues, corresponding to two linear
independent eigenvectors of the positive hermitian matrix $T_{\rho_{d\times
d}}^{^{\dag}}T_{\rho_{d\times d}},$ and $l_{d}=1$ if a qudit dimension
$d\geq2$ is even and $l_{d}=\sqrt{\frac{d-1}{d}}$ if a qudit dimension
$d\geq2$ is odd.
\end{theorem}

For each two-qubit state, the lower and upper bounds in (\ref{51}) coincide
and (\ref{51}) reduces to
\begin{equation}
\max_{A_{i},B_{k}\in\mathcal{L}_{2}}\text{ }\left\vert \mathrm{tr}%
[\rho_{2\times2}\mathcal{B}_{chsh}(A_{1},A_{2};B_{1},B_{2})]\right\vert \text{
}=\text{ }2\sqrt{\lambda_{\rho_{2\times2}}+\widetilde{\lambda}_{\rho
_{2\times2}}}, \label{51_1}%
\end{equation}
i. e. to the precise two-qubit result found by Horodeckis in \cite{9}.

As we prove in Section 4, for the two-qudit GHZ state, the upper bound
(\ref{51}) is also attained and gives the precise value for the maximum of the
CHSH expectation (\ref{35}) in this state.

\section{The two-qudit GHZ state}

Let us now specify the upper bound in (\ref{51}) for the two-qudit GHZ state
\begin{equation}
\rho_{ghz,d}=\frac{1}{d}\sum\limits_{j,k=1,...d}|j\rangle\left\langle
k\right\vert \otimes|j\rangle\left\langle k\right\vert . \label{54}%
\end{equation}
For this state, the correlation matrix $T_{\rho_{ghz,d}}$ is hermitian.
Calculating its elements due to relation (\ref{37}) and expressions (\ref{3}),
(\ref{4.1})--(\ref{4.3}), we come for this matrix to the following
diagonal-block form%
\begin{equation}%
\begin{pmatrix}
T^{(s)} & 0 & 0\\
0 & T^{(as)} & 0\\
0 & 0 & T^{(d)}%
\end{pmatrix}
\label{55}%
\end{equation}
where \newline(i) $T^{(s)}$ is the $\frac{d\left(  d-1\right)  }{2}\times
\frac{d\left(  d-1\right)  }{2}$ matrix with elements%
\begin{equation}
\mathrm{tr}[\rho_{ghz,d}\{\Lambda_{jk}^{(s)}\otimes\Lambda_{j_{1}k_{1}}%
^{(s)}\}],\text{ \ }\ j,j_{1},k,k_{1}=1,...,\frac{d(d-1)}{2}, \label{56}%
\end{equation}
which are equal to $\frac{2}{d}$ on the diagonal and to zero, otherwise;
\newline(ii) $T^{(as)}$ is the $\frac{d\left(  d-1\right)  }{2}\times
\frac{d\left(  d-1\right)  }{2}$ matrix with elements%
\begin{equation}
\mathrm{tr}[\rho_{ghz,d}\{\Lambda_{jk}^{(as)}\otimes\Lambda_{j_{1}k_{1}%
}^{(as)}\}],\text{ \ \ }\ j,j_{1},k,k_{1}=1,...,\frac{d(d-1)}{2}, \label{57}%
\end{equation}
which are equal to $(-\frac{2}{d})$ on the diagonal and to $0,$
otherwise;\newline(iii) $T^{(d)}$ is the $(d-1)\times(d-1)$ matrix with
elements%
\begin{equation}
\mathrm{tr}[\rho_{ghz,d}\{\Lambda_{l}^{(d)}\otimes\Lambda_{l_{1}}%
^{(d)}\}],\ \text{\ \ \ }l,l_{1}=1,...,d-1, \label{58}%
\end{equation}
which are equal to $\frac{2}{d}$ on the diagonal and to zero, otherwise.

Thus, for the two-qudit GHZ\ state (\ref{54}),
\begin{equation}
T_{\rho_{ghz,d}}^{\dag}T_{\rho_{ghz,d}}=T_{\rho_{ghz,d}}^{2}=\frac{4}{d^{2}%
}\text{ }\mathbb{I}_{\mathbb{R}^{d^{2}-1}}, \label{59}%
\end{equation}
and each vector $n\in\mathbb{R}^{d^{2}-1}$ is an eigenvector of $T_{\rho
_{ghz,d}}^{2}$ with eigenvalue $\frac{4}{d^{2}}.$ From (\ref{59}) it follows
that, for the GHZ state, the eigenvalues in (\ref{51})\ are given by
\begin{equation}
\lambda_{\rho_{ghz,d}}=\widetilde{\lambda}_{\rho_{ghz,d}}=\frac{4}{d^{2}},
\label{62}%
\end{equation}
so that, for the GHZ state, the general upper bound in (\ref{51}) reduces to%
\begin{equation}
\max_{A_{i},B_{k}\in\mathcal{L}_{d}}\text{ }\left\vert \text{ }\mathrm{tr}%
[\rho_{ghz,d}\mathcal{B}_{chsh}(A_{1},A_{2};B_{1},B_{2})]\text{ }\right\vert
\text{ }\leq\text{ }2l_{d}^{2}\sqrt{2}, \label{63}%
\end{equation}
where $l_{d}=1$ if a dimension $d\geq2$ is even and $l_{d}=\sqrt{\frac{d-1}%
{d}}$ if a dimension $d\geq2$ is odd. This proves the following statement.

\begin{proposition}
For the two-qudit GHZ state (\ref{54}) and traceless qudit observables with
eigenvalues in $[-1,1]$, the new upper bound introduced in Theorem 1 is equal
to $2l_{d}^{2}\sqrt{2}$ and, if a qudit dimension $d\geq2$ is odd, then this
upper bound is less than the general upper bound $2\sqrt{2}$ of Tsirelson
\cite{7,8}.
\end{proposition}

Furthermore, let us prove that, for the GHZ state, the upper bound in
(\ref{51}) is attained.

If $d=2$, then the two-qubit GHZ state constitutes one of the four states in
the Bell basis and the upper bound (\ref{63}) is attained.

Consider an arbitrary $d>2.$ From (\ref{55})--(\ref{58}) it follows that the
hermitian matrix $T_{\rho_{ghz,d}}$ has two proper subspaces $\mathfrak{J}%
_{\pm\frac{2}{d}}\subset\mathbb{R}^{d^{2}-1},$ corresponding to eigenvalues
$(\pm\frac{2}{d})$ and each vector $r\in\mathfrak{R}_{d}$ admits decomposition
$r=r^{(+)}+r^{(-)},$ where $r^{(\pm)}$ are projections of $r\in\mathfrak{R}%
_{d}$ onto the proper subspaces $\mathfrak{J}_{\pm\frac{2}{d}}$, respectively,
and
\begin{equation}
T_{\rho_{ghz,d}}r^{(\pm)}=\pm\frac{2}{d}r^{(\pm)},\text{ \ \ \ }\left\langle
r^{(+)},r^{(-)}\right\rangle =0. \label{64}%
\end{equation}
If projection $r^{(\pm)}\neq0,$ then it constitutes an eigenvector of
$T_{\rho_{ghz,d}}$ (not necessarily unit) corresponding to the eigenvalue
$\pm\frac{2}{d},$ respectively.

Let $X$ be a traceless qudit observable with: (i) $d$ mutually orthogonal unit
eigenvectors given by the unit vectors in the computational
basis\ $\{|j\rangle,$ $j=1,....,d\}$ of $\mathbb{C}^{d}$; (ii) eigenvalues
$\pm1$ if a dimension $d>2$ is even and eigenvalues $0,\pm1$ with multiplicity
$s=1$ of the zero eigenvalue -- if a dimension $d>2$ is odd.

By Definition 1, a traceless observable $X$ belongs to subset $\mathcal{L}%
_{d}^{(s)}\subset\mathcal{L}_{d}^{(s)},$ where $s=0$ if $d>2$ is even and
$s=1$ if $d>2$ is odd. The operator norm of this observable is equal to
$\left\Vert X\right\Vert _{0}=1.$Under representation (\ref{9}) to this
observable $X$ there corresponds vector $r_{X}\in\mathfrak{R}_{d}^{(s)},$
$s=0,1,$ where $\mathfrak{R}_{d}^{(s)}$ is given by (\ref{15}).

From (\ref{9}) it follows that components of vector $r_{X}$ are given by
$r_{X}^{(j)}=\frac{1}{\sqrt{2d}}\mathrm{tr}[X\Lambda_{j}],$ $j=1,....,(d^{2}%
-1).$ In view of (\ref{55}), (\ref{57}), components of $r_{X}$, corresponding
to projection $r_{X}^{(-)}$ onto the proper subspace $\mathfrak{J}_{-\frac
{2}{d}}\subset\mathbb{R}^{d^{2}-1}$ of $T_{\rho_{ghz,d}}$ corresponding to the
eigenvalue ($-\frac{2}{d}),$ are defined by traces $\frac{1}{\sqrt{2d}%
}\mathrm{tr}[X\Lambda_{jk}^{(as)}]$. Due to the structure (\ref{4.2}) of
operators $\Lambda_{jk}^{(as)}$ and the above specified structure of an
observable $X\in\mathcal{L}_{d}^{(s)},$ all these traces are equal to zero.
Therefore, $r_{X}^{(-)}=0$ and $r_{X}=r_{X}^{(+)}.$

Further, since $r_{X}=r_{X}^{(+)}\in\mathfrak{R}_{d}^{(s)},$ $s=0,1,$ we have
by (\ref{15}): $\left\Vert r_{X}^{(+)}\cdot\Lambda\right\Vert _{0}=\sqrt
{\frac{2}{d}}$ and $\left\Vert r_{X}^{(+)}\right\Vert =l_{d}$. Moreover, for
$d>2,$ we can always find at least two such observables $X_{1}$ and $X_{2},$
for which $r_{X_{1}}^{(+)}\neq-r_{X_{2}}^{(+)}.$

Taking all this into account for the maximum in the second line of (\ref{47}),
we derive:%
\begin{align}
&  \sqrt{2d}\max_{r_{1,}r_{2}\in\mathfrak{R}_{d}}\sqrt{\frac{\left\Vert
T_{\rho_{d\times d}}r_{1}\right\Vert ^{4}}{\left\Vert T_{\rho_{d\times d}%
}r\cdot\Lambda\right\Vert _{0}{}^{2}}+\frac{\left\Vert T_{\rho_{d\times d}%
}r_{2}\right\Vert ^{4}}{\left\Vert T_{\rho_{d\times d}}r_{2}\cdot
\Lambda\right\Vert _{0}^{2}}}\label{65}\\
& \nonumber\\
&  \geq\sqrt{2d}\left(  \sqrt{\frac{4}{d^{2}}\frac{\left\Vert r_{X_{1}}%
^{(+)}\right\Vert ^{4}}{\left\Vert r_{X_{1}}^{(+)}\cdot\Lambda\right\Vert
_{0}^{2}}+\frac{4}{d^{2}}\frac{\left\Vert r_{X_{2}}^{(+)}\right\Vert ^{4}%
}{\left\Vert r_{X_{2}}^{(+)}\cdot\Lambda\right\Vert ^{2}}}\right)  =2l_{d}%
^{2}\sqrt{2}.\nonumber
\end{align}
Therefore, Eqs. (\ref{47}), (\ref{65}) imply%
\begin{equation}
\max_{A_{i},B_{k}\in\mathcal{L}_{d}}\text{ }\left\vert \text{ }\mathrm{tr}%
[\rho_{ghz,d}\mathcal{B}_{chsh}(A_{1},A_{2};B_{1},B_{2})]\text{ }\right\vert
\text{ }\geq\text{ }2l_{d}^{2}\sqrt{2}. \label{66}%
\end{equation}
Comparing (\ref{63}) and (\ref{66}), we come to the following \emph{new
result.}

\begin{theorem}
For the two-qudit GHZ state $\rho_{ghz,d}$ $\ $with an arbitrary $d\geq2,$ the
upper bound in Theorem 1 is attained for each $d\geq2$ and specifies the
maximal value of the CHSH expectation (\ref{35}) in this state
\begin{equation}
\max_{A_{i},B_{k}\in\mathcal{L}_{d}}\text{ }\left\vert \text{ }\mathrm{tr}%
[\rho_{ghz,d}\mathcal{B}_{chsh}(A_{1},A_{2};B_{1},B_{2})]\text{ }\right\vert
\text{ }=\text{ }2l_{d}^{2}\sqrt{2}. \label{67}%
\end{equation}
Here, $l_{d}^{2}=1$ if a qudit dimension $d\geq2$ is even and $l_{d}%
=\sqrt{\frac{d-1}{d}}$ if a qudit dimension $d\geq2$ is odd.
\end{theorem}

This result for the maximal value of the CHSH expectation in the GHZ state can
be also derived if to substitute the correlation matrix (\ref{55}) directly
into maximum (\ref{40}).

\section{Conclusions}

In the present paper, we have formulated and proved the properties (Theorem 1)
of the generalized Gell-Mann representation for traceless qudit observables
with eigenvalues in $[-1,1]$ and studied via this representation the maximal
value of the CHSH expectation in a general two-qudit state with an arbitrary
qudit dimension $d\geq2.$

For the maximal value
\begin{equation}
\max_{A_{i},B_{k}\in\mathcal{L}_{d}}\text{ }\left\vert \text{ }\mathrm{tr}%
[\rho_{d\times d}\mathcal{B}_{chsh}(A_{1},A_{2};B_{1},B_{2})]\text{
}\right\vert \label{68}%
\end{equation}
of the CHSH expectation in a general two-qudit state $\rho_{d\times d},$
$d\geq2,$ and traceless qudit observables with eigenvalues in $[-1,1],$ we
have derived the precise expression (\ref{47}) via the correlation matrix
(\ref{37}) for this two-qudit state. This expression explicitly leads to the
upper bound of Tsirelson \cite{7,8}, and to \emph{two new bounds (\ref{51}),
lower and upper, }expressed\emph{ }(Theorem 2) via the spectral properties of
the correlation matrix for a two-qudit state $\rho_{d\times d},$ $d\geq2$.

We have not yet been able to specify if the new upper bound in (\ref{51})
improves the Tsirelson upper bound for each two-qudit state. However, this is
the case:\newline(i) for each two-qubit state, where the new lower bound and
the new upper bound coincide and reduce to the precise value of (\ref{68})
found by Horodeckis \cite{9};\smallskip\ \newline(ii) for the two-qudit GHZ
state (\ref{54}) with an arbitrary odd $d\geq2,$ where the new upper bound is
less (Proposition 1) than the upper bound of Tsirelson \cite{7,8}.\smallskip

Moreover, for the two-qudit GHZ\ state (\ref{54}), we have explicitly found
its correlation matrix (\ref{55}) and proved (Theorem 3) that, for the
two-qudit GHZ\ state with an arbitrary qudit dimension $d\geq2,$ the new upper
bound in (\ref{51})\ is attained and this specifies the following \emph{new
result: }for the GHZ state\emph{ }(\ref{54}),\emph{ }the maximum of the
CHSH\ expectation over traceless qudit observables with eigenvalues in
$[-1,1]$ is equal to $2\sqrt{2}$ if $d\geq2$ is even and to $\frac{2(d-1)}%
{d}\sqrt{2}$ if $d>2$ is odd.

\section{Appendix A}

Consider the proof of Lemma 1. The operator norm of a qudit observable
$n\cdot\Lambda$ is given by
\begin{equation}
\left\Vert n\cdot\Lambda\right\Vert _{0}:=\sup_{\left\Vert \psi\right\Vert
=1,\psi\in\mathbb{C}^{d}}\text{ }\left\vert \left\langle \psi,\left(
n\cdot\Lambda\right)  \psi\right\rangle \right\vert =\sup_{\left\Vert
\psi\right\Vert =1,\psi\in\mathbb{C}^{d}}\text{ }\left\vert \text{
}\mathrm{tr}[\left(  n\cdot\Lambda\right)  \left\vert \psi\right\rangle
\left\langle \psi\right\vert ]\text{ }\right\vert \tag{A1}\label{A1}%
\end{equation}
For a pure state $\left\vert \psi\right\rangle \left\langle \psi\right\vert ,$
the normalized version of decomposition (\ref{1}) reads
\begin{equation}
\left\vert \psi\right\rangle \left\langle \psi\right\vert =\frac
{\mathbb{I}_{\mathbb{C}^{d}}}{d}+\sqrt{\frac{d-1}{2d}}\left(  r_{\psi}%
\cdot\Lambda\right)  ,\text{ \ \ \ \ }r_{\psi}^{(j)}=\sqrt{\frac{d}{2(d-1)}%
}\text{ }\langle\psi\mathrm{,}\Lambda_{j}\psi\rangle]\text{\ } \tag{A2}%
\label{A2}%
\end{equation}
where $r_{\psi}\in\mathbb{R}^{d^{2}-1}.$ Since \textrm{tr}$[\Lambda_{j}]=0,$
it follows from (\ref{A2}) and (\ref{2}) that
\begin{equation}
1=\frac{1}{d}+\frac{d-1}{d}\left\Vert r_{\psi}\right\Vert ^{2}\text{
\ \ \ }\Leftrightarrow\text{ \ \ \ }\left\Vert r_{\psi}\right\Vert =1,\text{
\ \ }\left\Vert \psi\right\Vert =1,\text{ \ }\forall\psi\in\mathbb{C}^{d}.
\tag{A3}\label{A3}%
\end{equation}
Substituting (\ref{A2}) into (\ref{A1}) and taking into account (\ref{2}),
(\ref{A3}), we have%
\begin{equation}
\left\Vert n\cdot\Lambda\right\Vert _{0}=\sqrt{\frac{2(d-1)}{d}}\text{
}\left(  \sup_{\left\Vert \psi\right\Vert =1,\psi\in\mathbb{C}^{d}}\left\vert
\left\langle n,r_{\psi}\right\rangle \right\vert \right)  \text{ }\leq\text{
}\sqrt{\frac{2(d-1)}{d}}\left\Vert n\right\Vert . \tag{A4}\label{A4}%
\end{equation}
This proves the upper bound in (\ref{7}). To prove the lower bound and the
last upper bound in (\ref{7}), we use (\ref{6}) and relations
\begin{equation}
\left(  1+\delta_{d2}\right)  \left\Vert n\cdot\Lambda\right\Vert _{0}%
^{2}\text{ }\leq\text{ }\mathrm{tr}[(n\cdot\Lambda)^{2}]=2\left\Vert
n\right\Vert ^{2}\text{ }\leq\text{ }d\left\Vert n\cdot\Lambda\right\Vert
_{0}^{2}, \tag{A5}\label{A5}%
\end{equation}
which imply
\begin{equation}
\frac{2}{d}\text{ }\leq\text{ }\frac{\left\Vert n\cdot\Lambda\right\Vert
_{0}^{2}}{\left\Vert n\right\Vert ^{2}}\text{ }\leq\text{ }\frac{2}%
{1+\delta_{d2}}. \tag{A6}\label{A6}%
\end{equation}
Eqs. (\ref{A4}), (\ref{A6}) prove the statement of Lemma 1.

\section{Appendix B}

Consider the proof of Theorem 1. According to (\ref{23.1}) and (\ref{22})%

\begin{equation}
\sqrt{\frac{2}{d}}\text{ }\frac{\left\Vert T_{\rho_{d\times d}}n\right\Vert
}{\left\Vert T_{\rho_{d\times d}}n\cdot\Lambda\right\Vert _{0}}\leq
l_{d},\text{ \ \ }\forall n\in\mathbb{R}^{d^{2}-1}. \tag{B1}\label{B1}%
\end{equation}
Also, by the upper bound in (\ref{7})
\begin{equation}
\frac{\left\Vert T_{\rho_{d\times d}}n\right\Vert }{\left\Vert T_{\rho
_{d\times d}}n\cdot\Lambda\right\Vert _{0}}\text{ }\geq\text{ }\sqrt{\frac
{d}{2(d-1)}},\text{ \ \ }\forall n\in\mathbb{R}^{d^{2}-1}. \tag{B2}\label{B2}%
\end{equation}
Relations (\ref{B1}), (\ref{B2}) imply
\begin{equation}
\sqrt{\frac{d}{2(d-1)}}\text{ \ }\leq\text{ }\frac{\left\Vert T_{\rho_{d\times
d}}n\right\Vert }{\left\Vert T_{\rho_{d\times d}}n\cdot\Lambda\right\Vert
_{0}}\text{ }\leq\text{ }l_{d}\text{ }\sqrt{\frac{d}{2}} \tag{B3}\label{B3}%
\end{equation}
Substituting this into the maximum in the second line of (\ref{47}), we
derive
\begin{align}
&  \frac{d}{\sqrt{d-1}}\text{ }\max_{r_{1},r_{2}\in\mathfrak{R}_{d}}\text{
}\sqrt{\left\Vert T_{\rho_{d\times d}}r_{1}\right\Vert ^{2}+\left\Vert
T_{\rho_{d\times d}}r_{2}\right\Vert ^{2}}\tag{B4}\label{B4}\\
& \nonumber\\
&  \leq\sqrt{2d}\text{ }\max_{r_{1},r_{2}\in\mathfrak{R}_{d}}\text{ }%
\sqrt{\frac{\left\Vert T_{\rho_{d\times d}}r_{1}\right\Vert ^{4}}{\left\Vert
T_{\rho_{d\times d}}r_{1}\cdot\Lambda\right\Vert _{0}^{2}}+\frac{\left\Vert
T_{\rho_{d\times d}}r_{2}\right\Vert ^{4}}{\left\Vert T_{\rho_{d\times d}%
}r_{2}\cdot\Lambda\right\Vert _{0}^{2}}}\nonumber\\
& \nonumber\\
&  \leq l_{d}d\text{ }\max_{r_{1},r_{2}\in\mathfrak{R}_{d}}\text{ }%
\sqrt{\left\Vert T_{\rho_{d\times d}}r_{1}\right\Vert ^{2}+\left\Vert
T_{\rho_{d\times d}}r_{2}\right\Vert ^{2}}\nonumber
\end{align}
Taking further into account that, in view of (\ref{22}), (\ref{23}),
$\mathfrak{R}_{d}$ is a subset of the ball of radius $l_{d}$ and also contains
the ball of radius $\frac{1}{\sqrt{d-1}},$ we have
\begin{align}
&  \max_{r_{1,}r_{2}\in\mathfrak{R}_{d}}\text{ }\sqrt{\left\Vert
T_{\rho_{d\times d}}r_{1}\right\Vert ^{2}+\left\Vert T_{\rho_{d\times d}}%
r_{2}\right\Vert ^{2}}\tag{B5}\label{B5}\\
&  \leq\max_{\left\Vert r_{1}\right\Vert ,\left\Vert r_{2}\right\Vert \leq
l_{d}}\text{ }\sqrt{\left\Vert T_{\rho_{d\times d}}r_{1}\right\Vert
^{2}+\left\Vert T_{\rho_{d\times d}}r_{2}\right\Vert ^{2}}\nonumber\\
&  =\max_{\text{l.i.}\left\Vert r_{1}\right\Vert ,\left\Vert r_{2}\right\Vert
=l_{d}}\text{ }\sqrt{\left\Vert T_{\rho_{d\times d}}r_{1}\right\Vert
^{2}+\left\Vert T_{\rho_{d\times d}}r_{2}\right\Vert ^{2}}\nonumber
\end{align}
and
\begin{subequations}
\begin{align}
&  \max_{r_{1},r_{2}\in\mathfrak{R}_{d}}\text{ }\sqrt{\left\Vert
T_{\rho_{d\times d}}r_{1}\right\Vert ^{2}+\left\Vert T_{\rho_{d\times d}}%
r_{2}\right\Vert ^{2}}\tag{B6}\label{B6}\\
& \nonumber\\
&  \geq\max_{\left\Vert r_{1}\right\Vert ,\left\Vert r_{2}\right\Vert
\leq\frac{1}{\sqrt{d-1}}}\text{ }\sqrt{\left\Vert T_{\rho_{d\times d}}%
r_{1}\right\Vert ^{2}+\left\Vert T_{\rho_{d\times d}}r_{2}\right\Vert ^{2}%
}\nonumber\\
& \nonumber\\
&  =\max_{\text{l.i.}\left\Vert r_{1}\right\Vert ,\left\Vert r_{2}\right\Vert
=\frac{1}{\sqrt{d-1}}}\text{ }\sqrt{\left\Vert T_{\rho_{d\times d}}%
r_{1}\right\Vert ^{2}+\left\Vert T_{\rho_{d\times d}}r_{2}\right\Vert ^{2}%
}\nonumber
\end{align}
where abbreviation "\emph{l.i.}" in (\ref{B5}), (\ref{B6}),
means\emph{\ linear independent} and appears since the transition from
maximums in the second lines of (\ref{B5}), (\ref{B6}) to the maximums in the
third lines takes already into account maximums over vectors $r_{1},r_{2}$
along the same ray inside the ball, that is, linear dependent $r_{1},r_{2}$.

Therefore, from (\ref{B4})--(\ref{B6}) it follows
\end{subequations}
\begin{align}
&  \frac{d}{\sqrt{d-1}}\text{ }\max_{\text{l.i.}\left\Vert r_{1}\right\Vert
,\left\Vert r_{2}\right\Vert =\frac{1}{\sqrt{d-1}}}\text{ }\sqrt{\left\Vert
T_{\rho_{d\times d}}r_{1}\right\Vert ^{2}+\left\Vert T_{\rho_{d\times d}}%
r_{2}\right\Vert ^{2}}\tag{B7}\label{B7}\\
& \nonumber\\
&  \leq\sqrt{2d}\text{ }\max_{r_{1,}r_{2}\in\mathfrak{R}_{d}}\text{ }%
\sqrt{\frac{\left\Vert T_{\rho_{d\times d}}r_{1}\right\Vert ^{4}}{\left\Vert
T_{\rho_{d\times d}}r_{1}\cdot\Lambda\right\Vert _{0}^{2}}+\frac{\left\Vert
T_{\rho_{d\times d}}r_{2}\right\Vert ^{4}}{\left\Vert T_{\rho_{d\times d}%
}r_{2}\cdot\Lambda\right\Vert _{0}^{2}}}\nonumber\\
& \nonumber\\
&  \leq l_{d}d\text{ }\max_{\text{l.i.}\left\Vert r_{1}\right\Vert ,\left\Vert
r_{2}\right\Vert =l_{d}}\text{ }\sqrt{\left\Vert T_{\rho_{d\times d}}%
r_{1}\right\Vert ^{2}+\left\Vert T_{\rho_{d\times d}}r_{2}\right\Vert ^{2}%
}.\nonumber
\end{align}
Note also that, for each radius $R_{0}$ of the sphere in $\mathbb{R}^{d^{2}%
-1},$%
\begin{align}
&  \max_{\text{l.i.}\left\Vert r_{1}\right\Vert ,\left\Vert r_{2}\right\Vert
=R_{0}}\text{ }\sqrt{\left\Vert T_{\rho_{d\times d}}r_{1}\right\Vert
^{2}+\left\Vert T_{\rho_{d\times d}}r_{2}\right\Vert ^{2}}\tag{B8}\label{B8}\\
&  =R_{0}\sqrt{\lambda_{\rho_{d\times d}}+\widetilde{\lambda}_{\rho_{d\times
d}}},\nonumber
\end{align}
where $\lambda_{\rho_{d\times d}}\geq$ $\widetilde{\lambda}_{\rho_{d\times d}%
}\geq0$\textbf{\ }are two greater eigenvalues, corresponding to two linear
independent eigenvectors of the positive hermitian matrix $T_{\rho_{d\times
d}}^{^{\ast}}T_{\rho_{d\times d}}$.

Substituting (\ref{B8}) into (\ref{B7}), we derive%
\begin{align}
&  \frac{d}{d-1}\text{ }\sqrt{\lambda_{\rho_{d\times d}}+\widetilde{\lambda
}_{\rho_{d\times d}}}\tag{B9}\label{B9}\\
& \nonumber\\
&  \leq\sqrt{2d}\text{ }\max_{r_{1},r_{2}\in\mathfrak{R}_{d}}\text{ }%
\sqrt{\frac{\left\Vert T_{\rho_{d\times d}}r_{1}\right\Vert ^{4}}{\left\Vert
T_{\rho_{d\times d}}r_{1}\cdot\Lambda\right\Vert _{0}^{2}}+\frac{\left\Vert
T_{\rho_{d\times d}}r_{2}\right\Vert ^{4}}{\left\Vert T_{\rho_{d\times d}%
}r_{2}\cdot\Lambda\right\Vert _{0}^{2}}}\nonumber\\
& \nonumber\\
&  \leq l_{d}^{2}d\text{ }\sqrt{\lambda_{\rho_{d\times d}}+\widetilde{\lambda
}_{\rho_{d\times d}}}.\nonumber
\end{align}
In view of (\ref{47}), this proves the statement of Theorem 1.

\end{document}